\documentstyle[prl,aps,twocolumn,floats]{revtex}

\draft
\input psfig     
\def\be{\begin{equation}}
\def\ee{\end{equation}}

\def\mls{\delta}
\def\e{\varepsilon}

\begin{document}

\title{Universal gap fluctuations in the superconductor proximity
effect}

\author{M.~G.~Vavilov, P.~W.~Brouwer, and V.~Ambegaokar} 

\address{Laboratory of Atomic and Solid State Physics,
Cornell University, Ithaca, New York 14853}

\author{C.\ W.\ J.\ Beenakker} 

\address{Instituut-Lorentz, Universiteit Leiden,
P.O. Box 9506, 2300 RA Leiden, The Netherlands\\
{\small \rm (\today)}
~\\ \bigskip \parbox{14cm}{\small
%
Random-matrix theory is used to study the mesoscopic fluctuations of the 
excitation gap in a metal grain or quantum dot induced by the proximity to 
a superconductor. We propose that the probability distribution of the gap 
is a universal function in rescaled units. Our analytical prediction for 
the gap distribution agrees well with exact diagonalization of a model 
Hamiltonian.\medskip\\
PACS Numbers 73.23.-b, 74.50.+r, 74.80.Fp}}

\maketitle
\bigskip
A normal metal in the
proximity of a superconductor acquires characteristics that are
typical of the superconducting state \cite{prox}. 
One of those characteristics is that the quasiparticle density
of states vanishes at the Fermi energy. This superconductor
proximity effect is most pronounced in a confined geometry, such
as a thin metal film or metal grain, or a semiconductor quantum
dot. In that case, provided the scattering in the normal metal
is chaotic, no excitations exist within
an energy gap $E_{\rm g} \sim \hbar/\tau$, where $\tau$ 
is the typical time between collisions with the superconductor
\cite{Golubov,Belzig,MBFB,Lodder,Zhou,Ihra}.

If the coupling to the superconductor is weak (as for the point 
contact coupling of Fig.\ 1), the functional form of the density 
of states becomes independent of microscopic properties of the normal 
metal, such as the shape, dimensionality, or mean free path. Weak 
coupling means that $\tau$ is much bigger than the time $\tau_{\rm 
erg}$ needed for ergodic exploration of the phase space in the normal 
region. For a point contact with $N\gg 1$ propagating modes
at the Fermi level $\varepsilon = 0$, the density of states
has a square root singularity at the excitation gap
\cite{MBFB},
\begin{equation}
\label{1}
\rho_{\rm mf}(\e)=
  \frac{1}{\pi}\sqrt{\frac{\e-E_{\rm g}}{\Delta^3_{\rm g}}}.
\end{equation}
For a ballistic point
contact and in the absence of a magnetic field, $E_{\rm g} = c N \mls$
and $\Delta_{\rm g} = c' N^{1/3} \mls$, where $c = 0.048$ and
$c' = 0.068$ are numerical constants and $\mls$ is the mean level
spacing in the normal metal when it is decoupled from the
superconductor.

Equation (\ref{1}) was obtained in a self-consistent diagrammatic
perturbation theory that uses $\tau \mls/\hbar \sim N^{-1}$
as a small parameter. Such a mean-field
theory provides a smoothed density of states for
which energies can only be resolved 
on the scale of $\hbar/\tau \sim
N \mls$, not on smaller energy scales, and is unable to deal
with mesoscopic
sample-to-sample fluctuations of the excitation gap. 
Mesoscopic fluctuations arise, e.g., upon varying the shape of
a quantum dot or the impurity configuration in a metal grain.
The lowest excited state $\varepsilon_1$ fluctuates from sample
to sample around the mean field value $E_{\rm g}$, with a probability
distribution $P(\varepsilon_1)$. 
It is the purpose of this
paper to go beyond mean field theory and to study the mesoscopic
fluctuations of the excitation spectrum close to $E_{\rm g}$.
Our main result is that the gap distribution $P(\varepsilon_1)$ is a universal
function of the rescaled energy $x = (\varepsilon_1 - E_{\rm
g})/\Delta_{\rm g}$, 
in a broad range $|x| \ll N^{2/3}$. The Fermi level itself 
($\varepsilon=0$) falls outside this range, which is why the universal
gap distribution was not found in a recent related study \cite{NBA}.
Our main findings are illustrated in Fig.~2.

\begin{figure}
\centerline{\psfig{figure=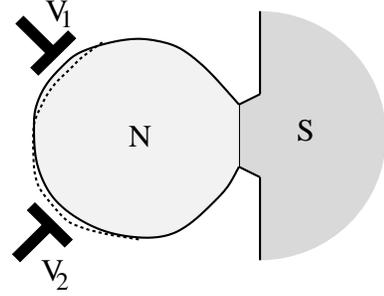,width=5cm}}
\bigskip
\narrowtext{
\caption{
A quantum dot (N) connected to a superconductor (S). 
The voltages on the gates $V_1$ and $V_2$ change the shape of the dot. 
Different values of the applied voltages create different samples
within the same ensemble.}
}
\end{figure}

We first consider the gap distribution in 
the absence of a magnetic field, and then include
a time-reversal symmetry breaking magnetic field.
Starting point of our calculation is the effective Hamiltonian 
\cite{FBMB}
\be
\label{3}
 {\cal H} = \left(
\begin{array}{cc}
 H & -\pi  W  W^\dagger \\
-\pi  W W^\dag & - H^*
\end{array}
\right).
\ee
Here $H$ is an $M\times M$ Hermitian matrix representing the
Hamiltonian of the isolated quantum dot, and $ W$ is an $M \times N$ 
matrix that describes the
coupling to the superconductor via an $N$-mode point contact.
For a ballistic point contact, $W_{mn} = \pi^{-1}
\delta_{mn} (M \mls)^{1/2}$ \cite{B}.
The number $M$ is sent to infinity at the end of the calculation
\cite{foot}.
The effective Hamiltonian is a valid description of the low-lying
excitations if the Thouless energy $N \mls$ is much smaller than the
order parameter $\Delta$ of the bulk superconductor. In the 
absence of a magnetic field, the matrix $H$ is symmetric. To describe
an ensemble of chaotic quantum dots (or disordered metal grains), we
take $H$ from the Gaussian orthogonal
ensemble (GOE) of random-matrix theory \cite{Mehta},
\be
\label{5}
{\cal P}(H)\propto \exp\left(-\frac{\pi^2}{4\mls^2M} {\rm Tr}{H}^2
\right).
\ee 
Calculation of the density of states 
of ${\cal H}$ using perturbation theory in $N^{-1}$
yields the result (\ref{1}) discussed in the introduction.
Our problem is
to go beyond perturbation theory and find the probability distribution
$P(\varepsilon_1)$ of the lowest positive eigenvalue $\varepsilon_1$
of the Hamiltonian (\ref{3}).

We have solved this problem numerically by exact diagonalization
of the effective Hamiltonian ${\cal H}$. Before presenting these
results, we first describe an entirely different approach, that
leads to an analytical prediction for the gap distribution. We
invoke the universality hypothesis of random-matrix theory, that
the local spectral statistics of a chaotic system depends only on
the symmetry properties of the Hamiltonian, and not on microscopic
properties. This universality hypothesis has been proven for a 
broad class of Hamiltonians in the bulk of the spectrum 
\cite{Guhr}, but is believed to be valid near the edge of the 
spectrum as well. A proof exists for so-called trace ensembles,
having ${\cal P}( H) \propto \exp[-\mbox{tr}\, f( H)]$, with
$f$ an arbitrary polynomial function \cite{Kanzieper}.

\begin{figure}
\centerline{\psfig{figure=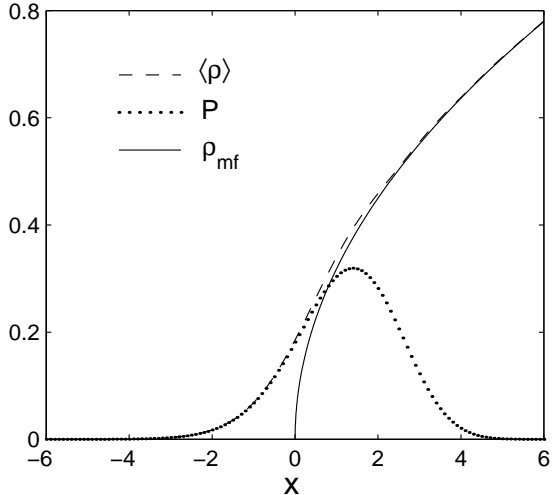,width=7.5cm}}
\bigskip
\narrowtext
{
\caption{
Mean field and ensemble averaged density of states $\rho_{\rm
mf}$ and $\langle \rho \rangle$, together with the probability 
distribution $P$ of the excitation gap,
as a function of the rescaled energy
$x=(\e_1-E_{\rm g})/\Delta_{\rm g}$.
These curves are the universal predictions of random-matrix theory.
}} 
\end{figure}

The mean-field density of states near the edge can be written in
the form
\be
  \rho_{\rm mf}(\varepsilon) = {1 \over a} 
  \left( {\varepsilon - b \over a} \right)^p,\ \
  \e > b.
\ee
According to the universality hypothesis, the spectral statistics
near the edge, in rescaled variables $(\varepsilon - b)/a$
depends only on the exponent $p$ and on the symmetry index
$\beta$ [$\beta=1$ ($2$) in the presence (absence) of time-reversal
symmetry]. Generically, $p$ is either $1/2$
(soft edge) or $-1/2$ (hard edge). For our problem, we have 
$\beta=1$, $p=1/2$, $a =  \pi^{2/3} \Delta_{\rm g}$,
$b=E_{\rm g}$, cf.\ Eq.\ (\ref{1}). The corresponding gap distribution
is given by \cite{TW}
\begin{eqnarray}
\label{8}
  P(\e) &=& \frac{d}{d\e}F_{1}\left[(\e-E_{\rm g})/\Delta_{\rm g}
\right], \\
\label{9}
  F_{1}(x) &=& \exp\left( -\case{1}{2} \int_{-\infty}^x [q(x')
  + (x - x') q^2(x')] dx' \right). 
\end{eqnarray}
The function $q(x)$ is the solution of
\be
\label{10}
q''(x)=-xq(x)+2q^3(x),
\ee
with asymptotic behavior
$q(x) \to {\rm Ai}(-x)$ as $x\to -\infty$ [${\rm Ai}(x)$ being
the Airy function].

The distribution (\ref{8}) is shown in Fig.~3 (solid curve). It is
centered at a positive value of 
$x = (\varepsilon_1 - E_{\rm g})/\Delta_{\rm g}$, meaning that the average
gapsize $\langle \e_1 \rangle$
is about $\Delta_{\rm g}$ bigger than the mean-field gap $E_{\rm g}$.
For small $x$ there is a tail of the form  
\begin{equation}
  P(x) \approx \frac{1}{4\sqrt{\pi}|x|^{1/4}}
  \exp \left( - \case{2}{3}|x|^{3/2} \right), \ \ \ x \ll - 1.  
\end{equation}
Non-universal corrections to the distribution (\ref{8}) become 
important for energy differences $|\e - E_{\rm g}| \gtrsim E_{\rm g}$, hence
for $|x| \gtrsim N^{2/3}$. Since the width of the gap distribution
is of order unity in the variable $x$, the probability to find a
sample with an excitation gap in the non-universal regime is
exponentially small.

In order to verify our universality hypothesis, we compare Eq.~(\ref{8}) 
with the results of an exact diagonalization of the
Hamiltonian (\ref{3}). As one can see in Fig.~3, the
numerical data are in good agreement with the analytical
prediction. The small deviations
can be attributed to the finiteness of $N$ and $M$
in the numerics.

Let us now consider the effect of a weak magnetic field on the
gap distribution. In the effective Hamiltonian, the presence of a 
magnetic field is
described by replacing $H$
by \cite{Pandey}
\begin{equation}
  H(\alpha) = H + i \alpha A. \label{PM}
\end{equation}
Here $A$ is an $M \times M$ real antisymmetric matrix, whose
off-diagonal elements have the same variance as those of $H$.
The parameter $\alpha$ is proportional to the magnetic
field,
\be
\label{13}
  M \alpha^2 = \eta \left(\frac{\Phi}{\Phi_0}\right)^2 
  {\hbar \over \tau_{\rm erg} \mls},
\ee
where $\Phi$ is the magnetic flux through the quantum dot,
$\Phi_0=h/e$ is the flux quantum, and $\eta$ is a non-universal
numerical
constant \cite{B}. The case
$\alpha=0$ corresponds to the GOE that we considered above; the
case $\alpha=1$ corresponds to the Gaussian unitary ensemble
(GUE) of fully broken time-reversal symmetry.

\begin{figure}
\centerline{\psfig{figure=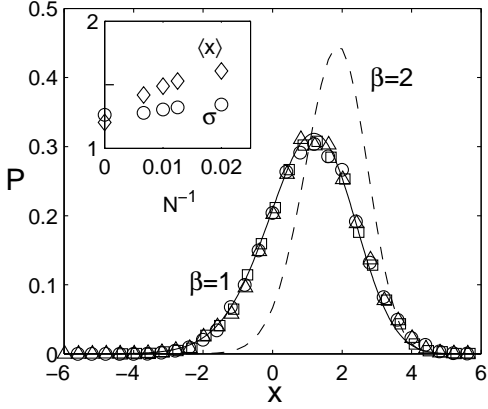,width=6.5cm}}
\narrowtext
{
\caption{
Probability distribution of the rescaled excitation gap $x =
(\varepsilon_1 - E_g)/\Delta_g$. 
Data points follow from an exact diagonalization of
$10^4$ realizations of the effective Hamiltonian (\protect\ref{3}) for
different values of $M$ and $N$ ($\bigtriangleup$: $M = 400$, $N=200$,
$\Box$: $M = 600$, $N = 150$, $\bigcirc$: $M = 600$,
$N = 80$). The solid curve is the universal
prediction (\protect\ref{8}) of random-matrix theory.
The mean of the data points has been adjusted to fit the curve by
applying an horizontal offset; no other fit parameters are involved.
The
inset shows the actual mean $\langle x \rangle$ and root-mean-square 
value $\sigma$ of the data for $M/N = 4$
for different values of $N$, together with the random-matrix
prediction for $N \to \infty$.
These results are all in zero magnetic field. 
The dashed curve is the random-matrix theory prediction 
(\protect\ref{P2}) in the presence of a time-reversal-symmetry
breaking magnetic field ($\beta=2$).
}}
\end{figure}

The effect of a magnetic field on the density of states in
mean-field theory is known \cite{MBFB}. The square-root
singularity (\ref{1}) near the gap still holds, but the 
magnitude of the gap is reduced. The critical flux $\Phi_{\rm c}$
at which $E_{\rm g}=0$ and hence the proximity effect is fully
suppressed is given by 
\be
  M \alpha^2 \sim N \ \Rightarrow \
  \Phi_{\rm c} \sim \Phi_0 \sqrt{\frac{N \tau_{\rm erg}  \mls}{\hbar}}.
\ee
This is a much larger flux than the flux $\Phi_{\rm bulk}$
at which the spectral statistics in the bulk of the spectrum
crosses over from GOE to GUE, which is given by \cite{Pandey}
\be
M \alpha^2 \sim 1 \ \Rightarrow \ \Phi_{\rm bulk} \sim
\Phi_0 \sqrt{\frac{\tau_{\rm erg} \mls}{\hbar}}.
\ee 

We will now argue that the characteristic flux $\Phi_{\rm edge}$ for
the spectral statistics at the edge of the spectrum is
intermediate between $\Phi_{\rm c}$ and $\Phi_{\rm bulk}$.
We consider the effect of the magnetic field 
on the lowest eigenvalue $\varepsilon_1$ of ${\cal H}$
to second order in perturbation theory,
\be
\label{15}
\delta \varepsilon_1=\sum_{j\neq 1}
  \alpha^2
  \frac{ |\langle 1| {\cal A} | j \rangle |^2}{\varepsilon_1-\varepsilon_j},
  \ \ \ \
  {\cal A} = 
  i \left( \begin{array}{cc} A & 0 \\ 0 & -A \end{array} \right).
\ee
Since typically $|\langle 1| {\cal A} | 2 \rangle |^2 \sim
M \mls^2/\pi^2$ and $\varepsilon_2 - \varepsilon_1
\sim \Delta_{\rm g}$, we see that the effect of level repulsion from the 
neighboring level $\varepsilon_2$ on the lowest level $\varepsilon_1$ 
becomes comparable to $\Delta_{\rm g} \sim N^{1/3} \mls$ if
\be
  M \alpha^2 \sim N^{2/3}\ \Rightarrow \
  \Phi_{\rm edge} \sim \Phi_0 
\sqrt{\frac{N^{2/3} \tau_{\rm erg}  \mls}{\hbar}}.
  \label{eq:PhiGap}
\ee
The terms in Eq.\ (\ref{15}) with $j \gg 1$ give
a uniform shift of all low-lying levels, and hence do not affect
the fluctuations.
For $N \gg 1$ the flux scale (\ref{eq:PhiGap}) for
breaking time-reversal symmetry at the edge of the spectrum is much
smaller than the critical flux $\Phi_{\rm c}$
needed to suppress the proximity effect. 
What is needed is $N^{2/3} \ll N$. 
This condition is difficult to satisfy in a numerical
calculation. The analytical prediction for
fully broken time-reversal symmetry is \cite{TW}
\begin{eqnarray}
  P(\varepsilon) &=& {d \over d\varepsilon} F_2[(\e - E_{\rm
  g})/\Delta_{\rm g}],
  \label{P2} 
  \\
  F_2(x) &=& \exp \left( - \int_{-\infty}^x (x - x') q^2(x') dx' \right).
\end{eqnarray}
This curve is shown dashed in Fig.\ 3. The tail for small $x$  
is now given by 
\begin{equation}
  P(x)  \approx \frac{1}{8\pi |x|} \exp\left(- \case{4}{3}|x|^{3/2}
\right), \ \ \ x \ll -1.
\end{equation}

To make contact with Ref.\ \onlinecite{NBA}
we briefly discuss the implications of 
our results for the ensemble averaged density of states 
$\langle\rho(\varepsilon)\rangle$ in the sub-gap regime. The tail of 
$P(x)$ for $x\lesssim -1$ is the same as the tail of 
$\langle\rho\rangle$, cf.\ Fig.\ 2. We conclude that \cite{foot2}
\be
  \langle \rho(x) \rangle \propto  \exp\left(-\frac{2\beta}{3}x^{3/2}\right)
\label{univdecay}
\ee
over a broad range $\Delta_{\rm g} \ll E_{\rm g} - \e \ll E_{\rm g}$
inside the mean--field gap. A
different exponential decay (with a power $2$ instead of $3/2$ in
the exponent) was predicted recently by Beloborodov, Narozhny, and
Aleiner \cite{NBA}, for the sub-gap density of states of an ensemble
of superconducting grains in a weak magnetic field. Since the
mean-field density of states in that problem is also of the
form (\ref{1}), the universal GUE edge statistics should apply.
The reason that the universal decay (\ref{univdecay}) was not
obtained in Ref.\ \onlinecite{NBA} is that their theory applies
to the non-universal energy range $\e \ll E_{\rm g}$ near the Fermi
level. To emphasize the significance of the universal energy range
we note that the probability to have the lowest energy level 
in that range is larger than in the non-universal range by an
exponentially large 
factor $\propto \exp[(E_{\rm g}/\Delta_{\rm g})^{3/2}]$.

In conclusion, we have argued that the proximity effect in a
mesoscopic system has a gap distribution which is universal
once energy is measured in
units of the energy scale $\Delta_{\rm g} \propto
(E_{\rm g} \mls^2)^{1/3}$ defined from the mean-field density of states
$\rho(\varepsilon) = [(\varepsilon-
E_{\rm g})/\Delta_{\rm g}^3]^{1/2}/\pi$. This universal distribution is
the same as 
the distribution of the smallest eigenvalue of the Gaussian
orthogonal or unitary ensembles from random-matrix theory, depending
on whether time-reversal symmetry is broken or not. We have 
identified the magnetic field scale for breaking time-reversal
symmetry and verified our results by exact diagonalization of an
effective Hamiltonian. Characteristic energy and magnetic field
scales are summarized in Table 1. The universality of our
prediction should offer ample opportunities for experimental
observation.

\begin{table}
\begin{tabular}{ c  c  c }
  & Energy scale   & Flux scale              \\
\hline
Bulk statistics & $\mls$ & $\Phi_0
  \tau_{\rm erg}^{1/2} \mls^{1/2}/\hbar^{1/2}$ \\
Edge statistics & $E_{\rm g}^{1/3} \mls^{2/3}$ & $\Phi_0 \tau_{\rm erg}^{1/2}
  \delta^{1/6} E_{\rm g}^{1/3}/\hbar^{1/2}$ \\
Gap size & $E_{\rm g}$ & $\Phi_0 \tau_{\rm erg}^{1/2} E_{\rm
g}^{1/2}/\hbar^{1/2}$ \\ 
\end{tabular}
\caption{
Characteristic energy and magnetic flux scales for the
spectral statistics in the bulk and at the edge of the spectrum 
and for the size of the gap.
}
\end{table}

We thank I.\ Aleiner, I.\ Beloborodov, E.\ Mishchenko,
and B.\ Narozhny for useful 
discussions. This work was supported by the Cornell Center for Materials 
Research under NSF grant No. DMR--9632275 and by the Dutch Science
Foundation NWO/FOM.


\begin{references}
\bibitem{tauerg}
In a quantum dot or metal grain of size $R$,
with Fermi velocity $v_{\rm F}$ and mean free path $\ell$,
one has $1/\tau_{\rm erg} \sim v_{\rm F} R^{-2} \min(\ell,R)$.
\bibitem{foot}
In our numerical computations it was necessary to choose the ratio
$M/N$ relatively small in order to achieve the limit
$N \gg 1$ needed for universality of the gap distribution. For 
finite $M/N$, the mean-field result (\ref{1}) still
holds, but now with coefficients $c$ and $c'$ for the energy scales
$E_{\rm g}$ and $\Delta_{\rm g}$ that weakly depend on $M/N$. For the
comparison 
of the numerical data with the prediction of random-matrix
theory we calculated $E_{\rm g}$
and $\Delta_{\rm g}$ from the mean-field theory (Eq.\ (8) of
Ref.\ \onlinecite{MBFB})
for the values of $M/N$ used in the exact diagonalizations.
No fit parameters are involved in this procedure.
\bibitem{foot2} The complete random matrix-theory prediction 
is $\langle \rho(x) \rangle
=-x{\rm Ai}^2(x)+[{\rm Ai}'(x)]^2 + \case{1}{2} \delta_{\beta,1} {\rm Ai}
(x)[1-\int_x^{\infty} {\rm Ai}(y)dy ]$. The $\beta=1$ result is
plotted in Fig. 2.
\end{references}
\end{document}